# HandiMathKey-Device

Frédéric Vella[1]; Nathalie Dubus[2], Eloise Grolleau[2], Marjorie Deleau[2], Cécile Malet[2], Christine Gallard[2], Véronique Ades[2] and Nadine Vigouroux[1]

[1] IRIT, UMR CNRS 5505, Paul Sabatier University, 118 Route de Narbonne, 31062 Toulouse Cedex 9, France
[2] ASEI, Jean Lagarde Center, 1 Avenue Tolosane, 31520 Ramonville-Saint-Agne, France
`frederic.vella@irit.fr`

**Abstract.** Typing mathematics is sometimes difficult with text editor functions for students with motor impairment and other associated impairments (visual, cognitive). Based on the HandiMathKey software keyboard, a user-centred design method involving the ecosytem of disabled students was applied to design the HMK-D physical keyboard for mathematical input. We opted for the Stream Deck device because of its multimedia features and its appeal to young students to the HMK-D. Preliminary tests with 8 students (5 in secondary school and 3 in high school) shows that HMK-D is highly accepted, accessible and fun for mathematical input by students with impairments. A longitudinal study of the usability and acceptability of HMK-D is planned for the 2023-2024 school year.

## 1  Introduction

Information and communication technologies on computers, or tablets, can become an assistive technology that makes the learning process more accessible. Indeed, handwriting is a difficult and tiring task for students with grapho-motor deficits. Benoit and Sagot [1] have analyzed and identified the difficulties encountered by students with neurodevelopmental disorders in order to determine special educational needs.

There is a poorly addressed input area in the accessibility field that deals with the input of scientific elements including mathematical formulas. For students, the ability to produce written work is an essential activity when they start secondary school. For those who have not acquired this autonomy in writing, they are accompanied by a secretary or by a companion of students with impairment who provides the written transcription. For others, the computer is an assistive tool used to compensate for handwriting that may be dysorthographic, slow and tiring. There are many text input solutions for motor impairment [2]. The work carried out during occupational therapy sessions enables the student to gradually acquire more functional typing, with or without the help of word prediction software with or without voice synthesis, and/or voice recognition software. In this way, students gradually develop autonomy in their schoolwork, being able to produce more independently when taking notes, doing homework and so on.

However, when it comes to typing mathematics, the computer is not an easy tool. Observations by occupational therapists at Centre for inclusive schooling Jean Lagarde of ASEI have shown that word-processor equation editors (Microsoft Office, Open Office, Libre Office, etc.) make it difficult for students with motor impairment to enter mathematical data. These include motor impairments (combining keys on the physical keyboard, bimanual coordination, multiple movements of the mouse to access symbols), visual and visuospatial difficulties (spatial location and localization of symbols in an environment overloaded with information and of small size, little contrast) and memory difficulties (memorizing shortcuts or writing codes).

Typing in these editors has proved to be demanding both functionally (motor impairment) and cognitively (attentional, visuo-spatial, memory), generating fatigue at every level for little productive and effective gain [3].

To avoid this fatigue when typing mathematical formulas, we co-designed the HMK mathematical input application [3]. The Centre for inclusive schooling Jean Lagarde of ASEI conducted a multidisciplinary workshop with 23 with different disabilities to evaluate its effectiveness and usefulness in three classes [4]. The use of the HMK (virtual HandiMathKey application) in the classroom favours its acceptability and appropriation. However many students identified the need for a physical HMK-D keyboard for those who have difficulty pointing.

In this paper, we will first look at the various existing solutions for entering mathematical formulae, then describe the genesis of HMK and report on the main conclusions drawn from its use. We will then detail the design process used to develop the HMK-D physical keyboard and describe the various prototypes produced. Finally, we will report on the initial feedback from observation of use by nine students with impairment.

## 2      Related work

Inputting mathematical formulae is a question of accessibility that can unfortunately be addressed.. According to Akpan and Beard [4] writing mathematical symbol is important in their learning. However, Smith [6] have shown that teaching mathematics with a computer-like digital artifact does not work well. In addition, handwriting is a difficult and tiring task for students with grapho-motor deficits [7]. Benoit & Sagot [1] have analyzed and identified the difficulties encountered by a student with neurodevelopmental disorders in order to determine special educational needs. This is why keystroking on the computer keyboard, combined with word processing software, is recommended for text entry.

Few studies have addressed this issue although Word and Open Office editors offer input interfaces consisting of button bars associated with mathematical symbols and an "input sheet". The analysis of input activity with these tools with disabled children has revealed that the use of these bars is complex and tiring. Bertrand *et al*. [3] have conducted an introspective study of some interactive applications (Dmaths [8], Math-

Type[9], MathMagic Lite [10], MathCast [11]) to create mathematical notation. Windsteiger [12] has designed a graphical user interface based on the possibility to have dynamic objects (sliders, menus, checkboxes, radio buttons, and more) but within the specific framework of the Mathematica programming environment. Their goal was to facilitate the use of the Mathematica programming environment. Elliott and Bilmes [13] proposed the CamMath application that allows the creation and manipulation of mathematical formulas using a speech recognition system. They reported that this input modality is useful for students or professionals with motor disabilities. In addition, the use of this modality results in fewer errors and faster input of mathematical formulas than when using a keyboard and pointing device [14]. Indeed, Anthony *et al*. have explored a multimodal input method combining handwriting and speech. Their hypothesis is that the multimodal input may enhance computer recognition and aid user cognition. They reported that novice users were indeed faster, more efficient and enjoyed the handwriting modality more than a standard keyboard and mouse mathematics interface, especially as equation length and complexity increased. Bouck et al. 15explored a developed computer-based voice input, speech output calculator for students with visual impairments. They reported positive perceptions of the calculator, particularly noting the independence.

However, although speech recognition is a useful modality for people with motor and visual impairment, it could be on the one hand intrusive in crowded environments (schools, etc.) and on the other hand, it would have degraded performance in noisy environments.

Bertrand *et al*. [3] developed the HMK application for typing the mathematical formula to reduce the fatigue reported by the same authors during the use of applications [8], [9], [10], [11], etc.. The HMK application was accessible for students with motor and speech impairments [4]. ElSheikh and Najdi 16 studied the use of special math hardware keyboard. Their study reported that the math keyboard supports well the goal of mathematic communication for learning mathematics.

This related work shows that some studies are looking into more efficient interaction methods, hardware solutions or input applications that are independent of editors.

## 3 HMK Background

### 3.1 Description of HandiMathKey (HMK)

HandiMathKey is an application for inputting mathematical formulae. It has been designed separately for lower and upper secondary schools. It was co-designed in collaboration with teachers and occupational therapists [1].
The HMK interface consists of three sub-keyboards (see **Fig. 1**):

- at the top, the Latin alphabet versus Greek sub-keyboards, which is used to enter utterances or responses of a mathematical exercise,
- in blue, the operators block and the numeric keypad common to all mathematical concepts

- and finally the mathematical concept sub-keyboard, accessible by tabs. They are four mathematical concepts (probability, trigonometry, functions and geometry).

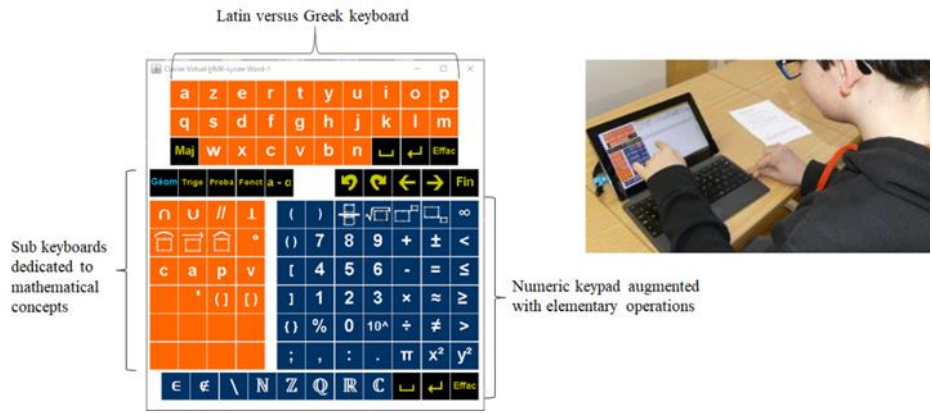

**Fig. 1.** HandiMathKey used by a student

### 3.2 Description of the HMK multidisciplinary study

A multidisciplinary team (a mathematics teacher, an occupational therapist and a special education assistant) led an interdisciplinary workshop to observe how students used HMK. This field study began in September 2018 at Centre for inclusive schooling Jean Lagarde of ASEI (Toulouse, France). HMK was pro-posed to a small group of students in three middle school classes (two 4th grade classes and one 5th grade class) during one school year. The study population consisted of 23 students: 19 with upper limb motor impairments, 3 with visual impairments and 1 with dyspraxia and dysgraphia. Every three weeks, the students used the HMK application during one hour of mathematics lessons. Each teacher taught both the mathematical concept and the associated mathematical symbols. The students' tasks consisted of copying mathematical formulae. Two classes used HMK with Microsoft Office, the third with Libre Office. We used written observations reports and a use log file of HMK as evaluation tools.

Observations of the three workshops [4] for each class confirmed the value of involving teams of mathematicians and occupational therapists in the HMK appropriation phase. The use of the HMK in the classes encouraged its acceptability and appropriation. No disability-related difficulties were reported. In all the classes, the students volunteered, but during the first two or three workshops they had not perceived any immediate interest in HMK. The students adopted HMK because of the easy access to mathematical symbols. However, this workshop demonstrated the need to first learn how to use the text editor to type mathematical formulae. Typing with HMK and Libre Office is similar to reading the mathematical formula, which makes it easier for students with planning and visual-spatial difficulties. .Using the HMK application with the Libre Office editor made it easier to learn its commands. Since in Microsoft Office the typing order is not similar to reading, typing with HMK application has to be planned and requires more visual attention on the part of the student.

This workshop revealed a new need to use a physical keyboard for the following reasons: difficulties with pointing, use of numeric operators out of routine. The aim of section 4 is to implement a user-centred design method for an HMK-D physical keyboard, based on the structuring of HMK into sub-keyboards.

## 4 User centered design

A user-centered co-design method [17] was implemented by three occupational therapists, two mathematics teachers, three computer science students and two senior researchers in human-computer interaction.

### 4.1 Expression of needs

During workshops on using the HMK application with lower secondary students, teachers and occupational therapists [4], it emerged that some students needed a physical HMK keyboard to enter mathematical formulae. Some of them had difficulty using the software application HMK with a pointing device. We also observed different input strategies: some used the physical keyboard to enter alphabetic characters for operators and numbers and the HMK application for the more complex mathematical symbols. These students expressed the need to have everything on one physical keyboard for mathematical input rather than having to juggle with the two input devices.

The device we had to design had to be USB-connected to the computer. It would either complement the physical keyboard-or replace it. We will refer to this solution as HMK-D in the remainder of this article.

### 4.2 Design cycles

Three co-design cycles were carried out, as described below. At each end of the design cycle, focus groups or user tests were carried out by occupational therapists and students with disabilities. These tests were carried out on several prototypes, first on very low-fidelity prototypes in the form of images, then on medium-fidelity prototypes implemented on physical keyboards and finally on high-fidelity prototypes implemented on a Stream desk.

**Low fidelity prototypes**
We planned to use two types of input device: a physical keyboard with mechanical buttons on which mathematical symbols would be associated, and a touch-sensitive tablet. The choice of the second medium was to free up the computer's field of vision so that only the text editor would be on the screen and the HMK keyboard on the entire tactile surface.

This expression of requirements led to six prototypes designed using a 3D modelling tool:

Proposal 1: A touch-sensitive tablet with an interface for inputting mathematic symbols. Above this is a zone for typing the mathematical formula. This provides visual continuity between the formula entered and the keys. Navigation between the interface is via tabs on the left of the interface (see

- **Fig. 2**);

Proposal 2: This is identical to proposal 1, except that navigation between the interfaces is via navigation arrows on either side of the interface (see

- **Fig. 3**);
- Proposal 3: A physical keyboard with mechanical keys containing mathematical symbols, the Greek and Latin alphabets and numeric operators. Above these keys, we have placed 3 switch keys enabling you to switch from one symbol to another for the same key (see **Fig. 4**); for example, the orange switch key enables you to switch from the Latin character block to the Greek symbol block.
- Proposal 4: Proposal 3 without the alphabetic keys (see **Fig. 5**) ;
- Proposal 5: Proposal 1 without the mathematical formula input area (see **Fig. 6**);
- Proposal 6: Proposal 2 without the mathematical formula input area (see **Fig. 7**).

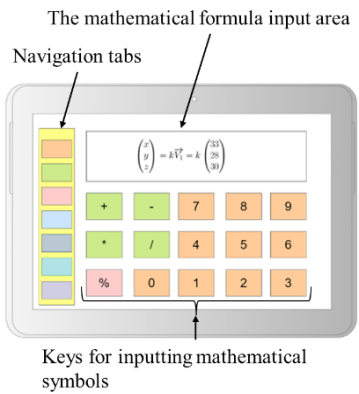
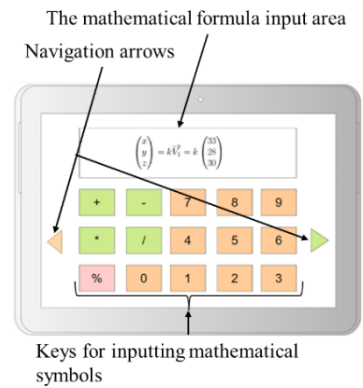

**Fig. 2.** Proposal 1 on touchpad.   **Fig. 3.** Proposal 2 on touchpad

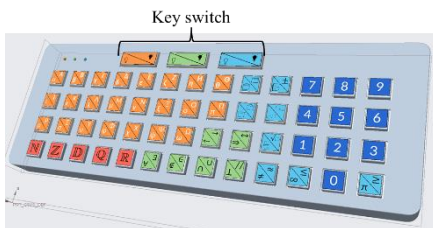
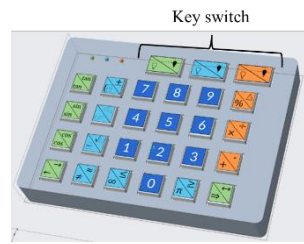

**Fig. 4.** Proposal 3   **Fig. 5.** Proposal 4

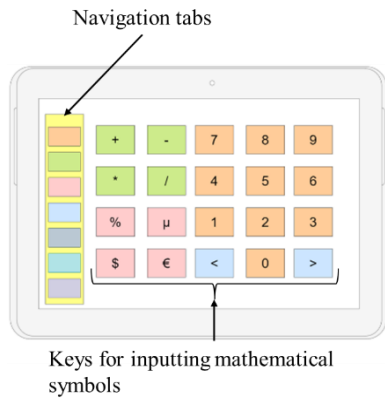 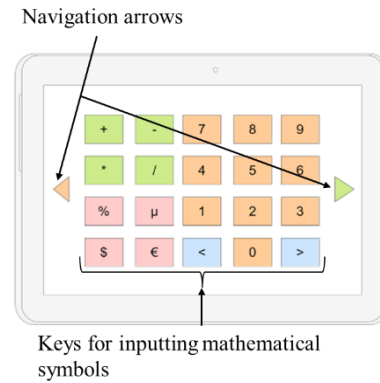

**Fig. 6.** Proposal 5 on touchpad    **Fig. 7.** Proposal 6 on touchpad

The chosen solution is proposal 3. This is the most complete solution compared with the HMK software version. Proposals 1, 2, 5 and 6 were not retained because of the high risk of occlusion during tactile interaction for children with neuromotor disorders of the upper limbs. The coexistence of the two keyboards (standard physical keyboard and solution 3 keyboard) means that the numeric keypad, for example, is redundant. In addition, the use of two large keyboards is cumbersome and generates significant motor movements. These drawbacks led us to design a prototype on an existing keyboard.

### Co-design cycles for a physical keyboard

The focus group conducted during the previous cycle/step led us to the design of a physical keyboard with mechanical keys. However, to avoid manufacturing costs and implementation time, we preferred to start with a conventional physical keyboard. In this cycle, we will successively present the mock-ups in paper format to be more representative of reality in relation to the size scale and then the physical solution of the paper mock-up.

*Paper prototypes*

In **Fig. 8** the static symbols (operators, numbers, symbols common to the four concepts, etc.) have been placed around the alphabet block. The numeric keypad, on the other hand, forms the dynamic part and has been replaced here by the symbols of the mathematical concept of trigonometry.

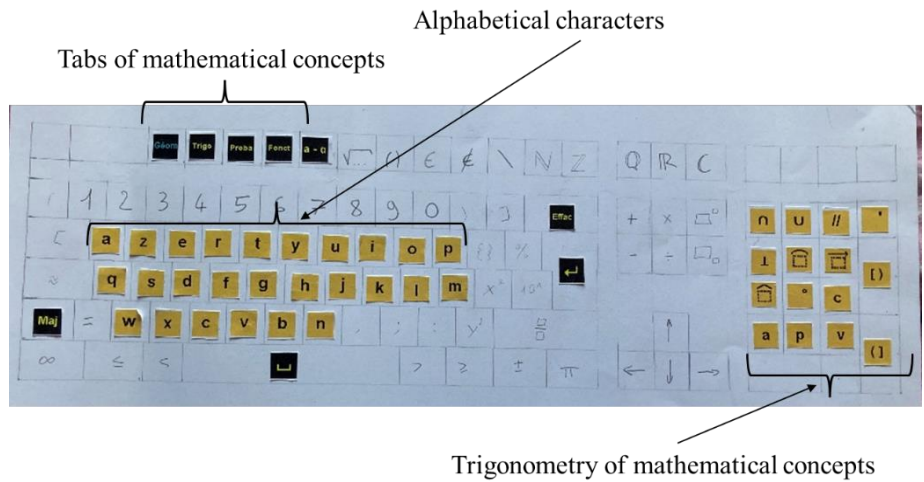

**Fig. 8.** Paper mock-up version 1

In **Fig. 9**, the symbols of the trigonometry concept have been arranged above the alphabet block instead of the numbers.

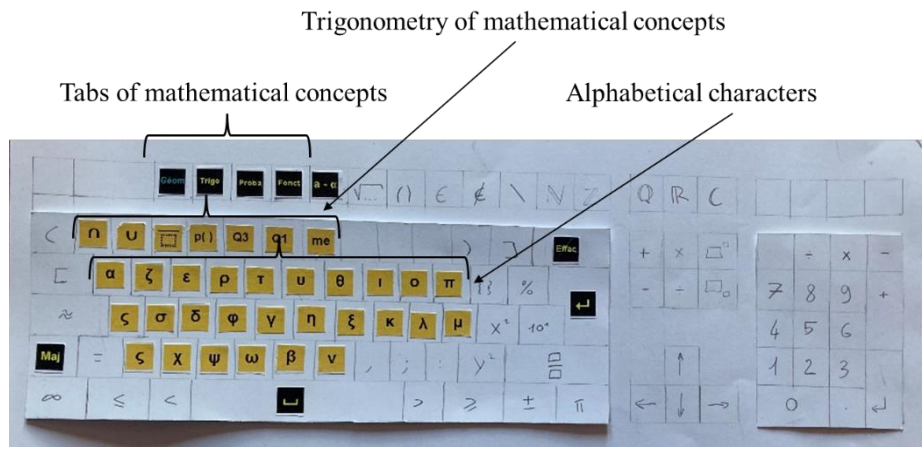

**Fig. 9.** Paper mock-up version 2

The most difficult part of these two mock-ups was defining the position of the 4 mathematical concepts. The geometry concept was either placed where the numeric keypad was (see **Fig. 8**), or where the numbers were located above the alphabetical section (see **Fig. 9**). This was not a good choice because of the children's acquired habits regarding the layout of the keyboard. At the end of the focus group between the occupational therapists and the human computer interaction researchers, it was suggested that all the static keys should be positioned on the physical keyboard and the dynamic part relating to the four mathematical concepts on another USB external port (see **Fig. 10**). We did not choose the solution of associating the 4 symbols with the same key because of the small size of the symbol's spelling, the additional attention and visual effort required to identify the right symbol and the complexity of the symbol selection mechanism (e.g. combining keys).

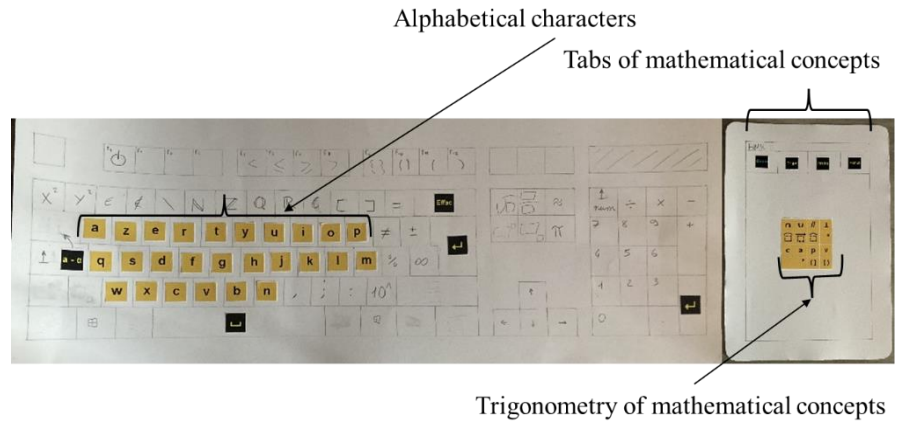

**Fig. 10.** Paper mock-up version 3.

*Physical prototypes*

A focus group with occupational therapists and mathematics teachers led to the following proposal (see **Fig. 11**) for the representation of mathematical formulae in white. A transducer based on Teensy version 4.1[1] converts the keys labelled white on the keyboard into mathematical symbols. These symbols are common to all mathematical concepts.

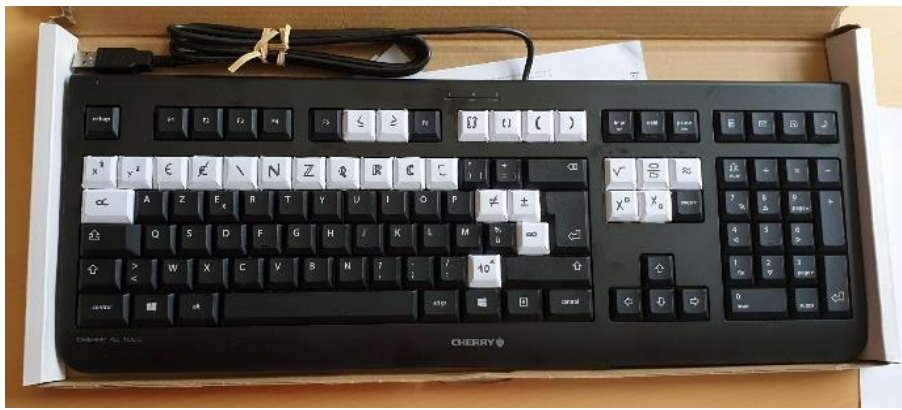

**Fig. 11.** Labelled physical keyboard

---

[1] https://www.pjrc.com/teensy/

As the dynamic part was not designated at this level, the occupational therapists only evaluated the static part with students. This first version of the HMK-D keyboard was acceptable to the children. After these user tests, the occupational therapists again asked for a dynamic part corresponding to the four mathematical concepts. They strongly emphasized that the solution had to be integrated to limit the number of devices (ergonomic improvements: cluttering up the student's desk and cumbersome installation of the two devices, etc.), fun and playful. We explored the technologies used in video games and multimedia. They also said that the characteristics (color or background) of the buttons should change according to the mathematical concept.

**Stream Deck prototype**

We selected the stream deck of the company Elgato© composed of 32 fully customizable LCD (Liquid Crystal Display) keys (see **Fig. 12**). For each keys it is possible to design its content (textual, sound, running a media, visual and sound feedback…). The first version of HMK-D on the Stream Deck has been designed as closely as possible to the sub-keyboard structure of the HMK software application (see **Fig. 1**). The tests were carried out using Microsoft.

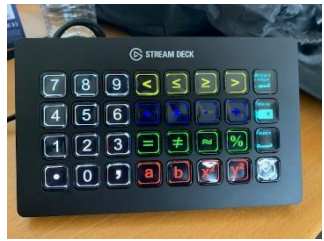

**Fig. 12.** Stream Deck

Together with the therapists, we produced several prototypes of the HMK-D keyboard on the Stream Deck following trials by disabled students from the Jean Lagarde Centre for Inclusive Education in secondary school (fifth and fourth) and high school (from seconde to terminale).

A page corresponds to a Stream Deck interface change. Pages are navigated by means of the buttons illustrated in (**Fig. 13**).

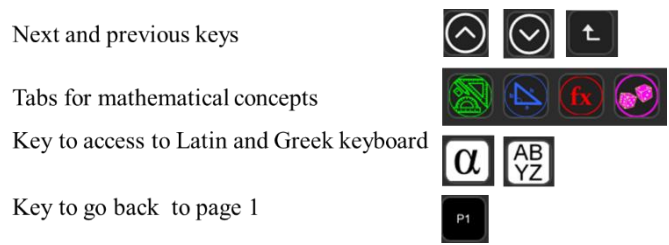

**Fig. 13.** Page change keys

We present below the changes per page that we have made to HMK-D from the first version that was tested (noted 'a' in the figures) to the latest version (noted 'b' in the figures). These changes were introduced following the consensus reached at focus group meetings between teachers, researchers and occupational therapists. One of the changes from page two was to add a button called P1 for returning to page one, to facilitate rapid access to operators.

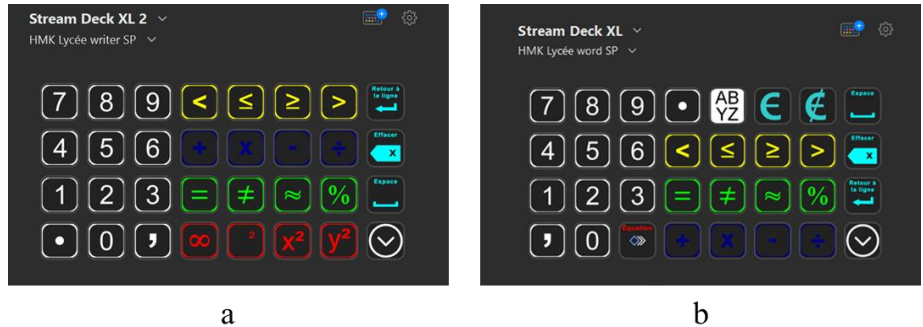

**Fig. 14.** Page 1

Page 1 consisted of the numeric keypad, operators and functionalities (see **Fig. 14** a). User tests identified the need to add the membership symbols, the alphabetic keyboard access key and the equations key (see **Fig. 14** b). This layout makes it easier to enter basic mathematical formulas.

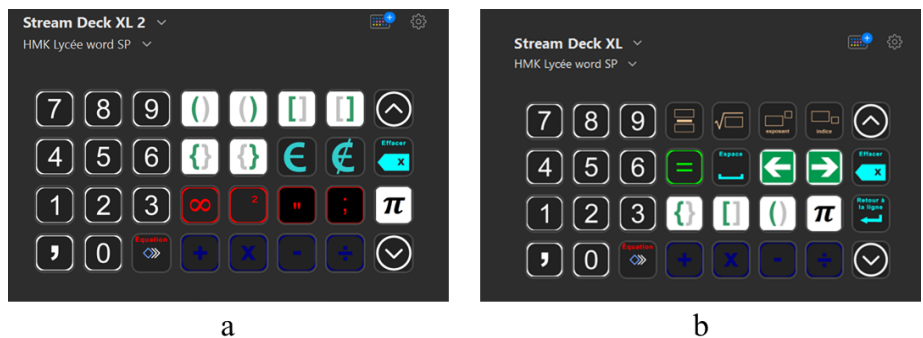

**Fig. 15.** Page 2

Page 2 has undergone a number of changes. In fact, this first version (see **Fig. 15** a) required a lot of navigating through the pages due to the incorrect structuring of HMK-D and the arrangement of symbols within the pages, as revealed during testing with children. So, in the latest version (see **Fig. 15**

**Fig. 15** b), we prefer to integrate the common functions (fraction, square root, etc.) with all the mathematical concepts by integrating the arrows for navigating formulas, spacing, the equal key and parenthesis.

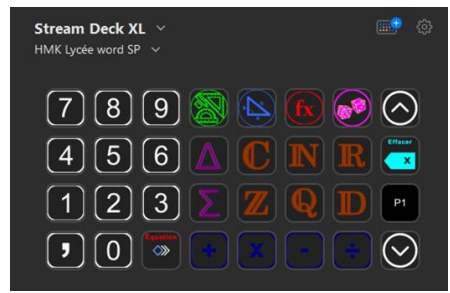

**Fig. 16.** Page 3

No changes have been made to page 3 (see **Fig. 16**). However, the pages of the mathematical concepts have been modified by adding navigation arrows in the formula to simplify its input. For example, on the page related to the mathematical function concept, the navigation arrows have been added to the bottom right-hand section (see **Fig. 17** b).

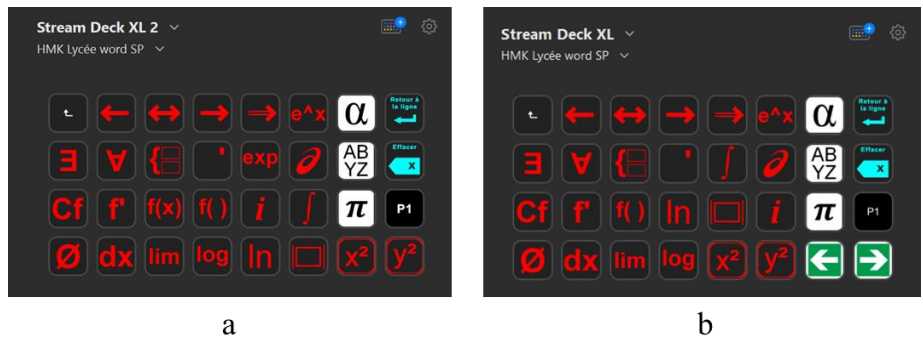

a                              b
**Fig. 17.** Page function concept

Page 4 was initially used to navigate mathematical formulas and to add the rest of the symbols (see **Fig. 18** a). The modifications made concerned the addition of the symbols ";" and the quotation mark, and a rearrangement of the operators. The navigation arrow were not considered useful on this page.

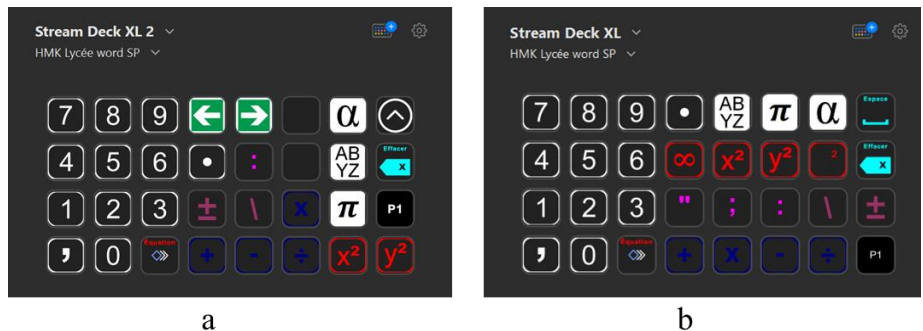

**Fig. 18.** Page 4

The occupational therapists had the HMK-D solution implemented on the Stream Desk empirically tested by 9 students of middle school and high school from the Centre for Inclusive Schools Jean Lagarde. These user experience was done during the rehabilitation session.

## 5 Preliminary results and perspective

8 students (5 in secondary school and 3 in high school) took part in tests using the HMK adaptation on the Stream Deck during the occupational therapy session at . 7 students have a motor disability and one has a neurodevelopmental disorder. 3 of them have associated learning disorders and two others have visual-spatial disorders. All 8 used a physical azerty keyboard. 7 of them had already used the HMK software version [3]. For these user tests we used a Thing Aloud type method [18] as the students expressed their feelings and comments verbally. The HMK version designed for the Stream Deck was generally accepted by the students. They also appreciated the visual feedback following the press of a key, the absence of noise unlike the HMK software version where click noise is present all the time and the possibility of tilting the Stream Deck which facilitates the spatial layout of the student's working environment (see **Fig. 19**).

The verbatims collected attest to the degree of satisfaction of the students: "*It's too good. I was tired with HMK and now it's better*", "*It's easy to learn*", grade 4 student. "*It's very good and better than HMK*", students from second. "*It's satisfying to press*", another grade 4 student. "*It's good and practical*", final year student. Stream Deck is also fun. "*It's more fun and easier*", a grade 4 student who is a regular user of the HMK application. Some students expressed an interest in using it in class. "*I'd prefer Stream Deck HMK-D for the classroom*", grade 4 student. "*I'm motivated to learn maths*", grade 4 student.

The Stream Deck's multimedia features enabled the HMK-D application to be designed in a more ergonomic, attractive and customisable way, and it was well accepted by the students during the series of pre-tests.

The first feedback from HMK-D users has been very promising. However, there are still improvements to be considered: the possibility of configuring the backlighting to take account of glare (students with visual-spatial disorders); the integration of an azerty keyboard rather than an alphabetical keyboard on the page for text input; the addition of character sound associated to the key. We also propose to make better use of the user profile function available in the Stream Deck environment to offer several profiles depending on the impairment of the students.

A multidisciplinary longitudinal study over the school year (empirical observations, HMK-D usage logs, usability questionnaires) on secondary school classes is planned for the start of the 2023-2024 school year on a population of pupils with motor impairment with or without associated other disorders.

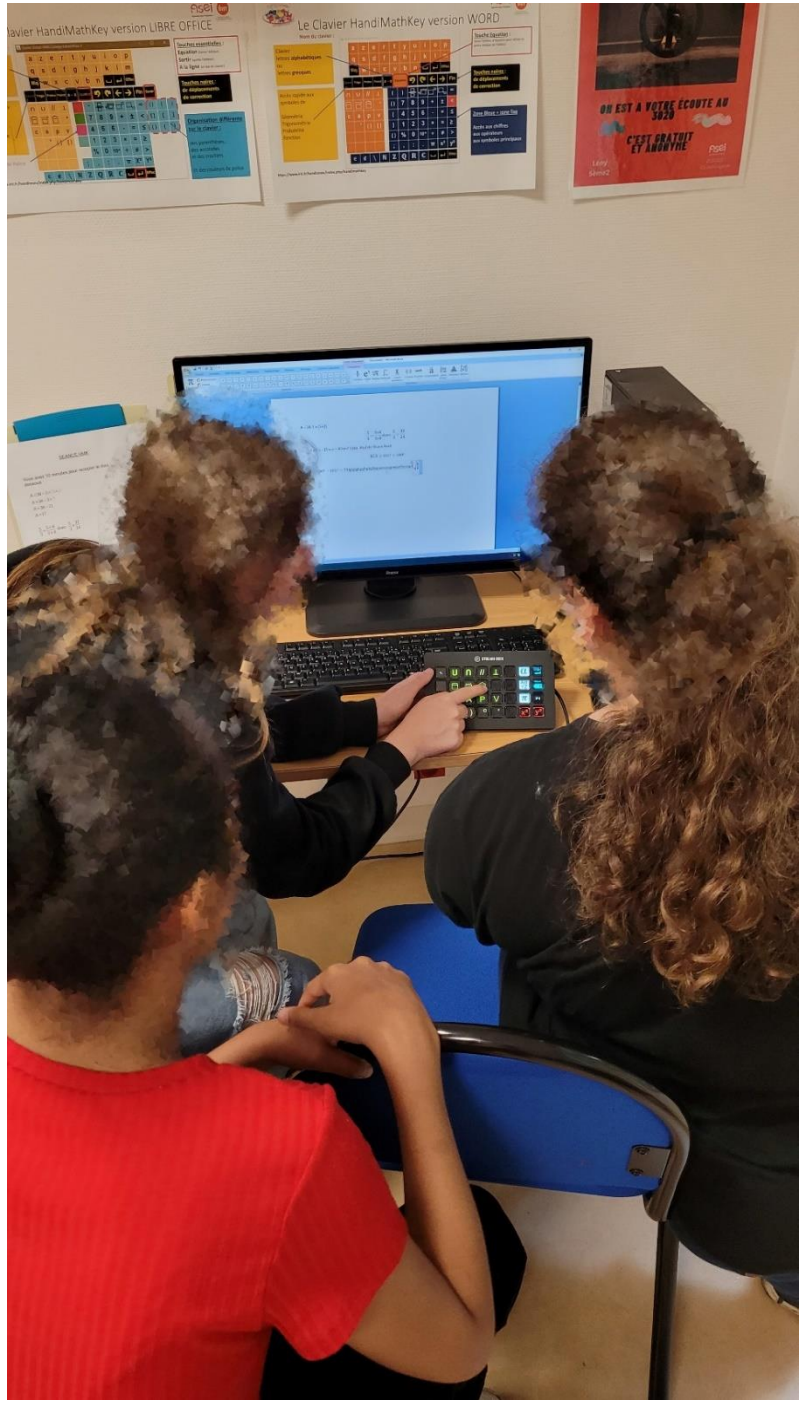

**Fig. 19.** Students using HMK-D on the Stream Deck

## 6      Conclusion

We used a user-centered design method to design the HMK-D mathematical input solution for secondary school students with impairment at Centre for inclusive schooling Jean Lagarde of ASEI. This co-design was carried out with occupational therapists, mathematics teachers and researchers and students in human-computer interaction.

Several HMK-D co-design cycles were implemented. Cycle 2 proposed a physical keyboard, but this was not chosen because of the size of the solution due to the addition of an extra device to implement the sub-keyboards corresponding to the mathematical concepts. We opted for the Stream Deck device because of its multimedia features and its appeal to young students. Initial feedback shows that HMK-D is highly accepted and accessible for mathematical input by students with impairments. A longitudinal study of the usability and acceptability of HMK-D is planned for the 2023-2024 school year.

### Acknowledgement

We would particularly like to thank the Centre for inclusive schooling Jean Lagarde of ASEI f(Toulouse, France) or enabling us to implement the co-design method as part of the Hand'Innov agreement between the IRIT laboratory and ASEI. We would also like to thank all the students with impairment who took part in the first user experiments with HMK-D.